\documentclass[reprint,superscriptaddress,amsmath,amssymb,aps,prd]{revtex4-1}

\usepackage{graphicx}
\usepackage{natbib}
\usepackage{hyperref}
\usepackage{cleveref}
\usepackage[caption=false,labelformat=empty]{subfig}
\usepackage{mathrsfs}
\usepackage{gensymb}

\hypersetup{colorlinks=true,citecolor=blue}

\newcommand{\Wcmsqd}{\mathrm{W}\text{cm}^{-2}}
\newcommand{\micron}{{\mu\mathrm{m}}}
\newcommand{\Power}{\mathcal{P}}

\newcommand{\Vm}{\text{V}\text{m}^{-1}}
\newcommand{\Ecrit}{E_\text{cr}}
\newcommand{\chimax}{\chi_0}
\newcommand{\neff}{n_\text{eff}}
\newcommand{\stdwidth}{\linewidth}

\begin{document}

\title{Reaching supercritical field strengths with intense lasers}

\author{T. G. Blackburn}
\email{tom.blackburn@physics.gu.se}
\altaffiliation[Present address: ]{Department of Physics, University of Gothenburg, SE-41296 Gothenburg, Sweden}
\affiliation{Department of Physics, Chalmers University of Technology, SE-41296 Gothenburg, Sweden}
\author{A. Ilderton}
\affiliation{Centre for Mathematical Sciences, Plymouth University, Devon PL4 8AA, United Kingdom}
\author{M. Marklund}
\altaffiliation[Present address: ]{Department of Physics, University of Gothenburg, SE-41296 Gothenburg, Sweden}
\affiliation{Department of Physics, Chalmers University of Technology, SE-41296 Gothenburg, Sweden}
\author{C. P. Ridgers}
\affiliation{York Plasma Institute, Department of Physics, University of York, York, YO10 5DD, United Kingdom}

\date{\today}

\begin{abstract}
It is conjectured that all perturbative approaches to quantum electrodynamics (QED) break down in the collision of a high-energy electron beam with an intense laser, when the laser fields are boosted to `supercritical' strengths far greater than the critical field of QED.
As field strengths increase toward this regime, cascades of photon emission and electron-positron pair creation are expected, as well as the onset of substantial radiative corrections.
Here we identify the important role played by the collision angle in mitigating energy losses to photon emission that would otherwise prevent the electrons reaching the supercritical regime.
We show that a collision between an electron beam with energy in the tens of GeV and a laser pulse of intensity $10^{24}~\Wcmsqd$ at oblique, or even normal, incidence is a viable platform for studying the breakdown of perturbative strong-field QED.
Our results have implications for the design of near-term experiments as they predict that certain quantum effects are enhanced at oblique incidence.
\end{abstract}

\maketitle

\section{Introduction}

Experimental exploration of nonperturbative quantum electrodynamics (QED)
is challenging as it requires electromagnetic fields comparable in strength
to the critical field of QED $\Ecrit =1.3\times10^{18}~\text{V}\text{m}^{-1}$,
the field strength which induces electron-positron pair creation from the vacuum
itself~\cite{sauter,schwinger.pr.1951}.
Nevertheless, ever-increasing laser intensities~\cite{mourou.rmp.2006,marklund.rmp.2006,dipiazza.rmp.2012}
make it possible to probe fields that are effectively supercritical, i.e., that have magnitude greater than $\Ecrit$.
This is achieved using the Lorentz boost when ultrarelativistic electrons collide
with an intense laser pulse~\cite{bula.prl.1996,burke.prl.1997},
as the parameter $\chi_e$ that controls the importance of nonlinear QED is
the rest-frame electric field normalized to the critical field strength $\Ecrit = m^2/e$.
$\chi_e$ is covariantly expressed as $\chi_e = |F_{\mu\nu} u^{\nu}|/\Ecrit$~\cite{ritus},
where $F$ is the electromagnetic field tensor and $u$ the electron four-velocity.
We use natural units in which $\hbar = c = 1$ ($e$ is the elementary charge,
$m$ the electron mass) throughout.

In the supercritical regime $\chi_e \gg 1$, particle dynamics is dominated by
cascades of photon emission and electron-positron pair creation%
~\cite{erber.rmp.1966,ritus,sokolov.prl.2010,bulanov.pra.2013}.
The importance of studying these phenomena is motivated by their relevance to
high-field astrophysical environments, such as magnetars%
~\cite{harding.rpp.2006,ruffini.prep.2010,timokhin.apj.2015},
and to laser-matter interactions beyond the current intensity frontier~\cite{bell.kirk,ridgers.prl.2012}.
It is also of considerable theoretical interest, as when $\alpha \chi_e^{2/3}$
approaches unity ($\alpha$ is the fine-structure constant),
it is conjectured that
radiative corrections to quantum processes become so large that
all current, perturbative, predictions fail~\cite{narozhny.prd.1980,fedotov.jpc.2017}
and strong-field QED becomes fully nonperturbative.

In this article we show how the collision of an intense laser pulse with
an ultrarelativistic electron beam may be used to probe the supercritical regime.
A significant obstacle to this is posed by radiation reaction,
an accelerating charge's loss of energy to photon emission, which strongly
reduces $u$ at $\chi_e \gg 1$, thereby suppressing $\chi_e$ itself%
~\cite{bulanov.ppr.2004,koga.pop.2005,hadad.prd.2010,kravets.pre.2013}.
We show that this can be mitigated by appropriate choice of the angle
at which the beams collide.
We present a theoretical expression for the maximum $\chi_e$,
which predicts,
contrary to the expectation that the ideal geometry is counterpropagation,
that oblique incidence is favoured for laser pulses of
high intensity or long duration.
This enhances certain quantum effects on radiation reaction,
namely straggling~\cite{shen.prl.1972,blackburn.prl.2014} and stochastic broadening~\cite{neitz.prl.2013}.
As a result, not only will laser-electron collision
experiments that are practically constrained to oblique incidence~\cite{eli.tdr}
still detect signatures of quantum effects, but these signatures
can be stronger than they would be in a head-on collision.
Furthermore, we show that at the intensity and electron energy necessary to probe
radiative corrections,
oblique, or even normal, incidence is strongly favoured to reduce radiative losses that
would otherwise prevent reaching such high $\chi_e$.

\section{Effect of radiative losses on the maximum $\chi_e$}

High-power lasers compress energy into ultrashort
pulses that can be focussed almost to the diffraction limit.
The theoretical upper bound on $\chi_e$ is obtained by treating the laser
as a pulsed plane electromagnetic wave with peak dimensionless
amplitude $a_0 = e E_0 / (m \omega_0)$, peak electric field strength $E_0$
and central frequency $\omega_0$, and neglecting radiative losses.
Then
	\begin{equation}
	\chi_e
		= \frac{a_0 \gamma_0 \omega_0 (1 + \cos\theta)}{m}
	\label{eq:ChiMax}
	\end{equation}
where $\theta$ is the collision angle (defined to be zero for
counterpropagation) and $\gamma_0 \gg 1$ is the initial Lorentz
factor of the electron.
The largest possible quantum parameter is $\chi_0 = \chi_e(\theta = 0)$.

Experiments at $a_0 \simeq 0.4$, $\chi_e \simeq 0.3$ have demonstrated
nonlinear QED effects including pair creation~\cite{bula.prl.1996,burke.prl.1997},
and recently evidence of radiation reaction has been
reported at $a_0 \simeq 10$, $\chi_e \simeq 0.1$~\cite{cole.prx.2018,poder.prx.2018}.
At present, the highest field strengths are equivalent to
$a_0 \simeq 50$~\cite{bahk,pirozhkov}, or $\chi_0 \simeq 1$ at $\gamma_0 m \simeq 1~$GeV;
$a_0 > 100$ is expected in the next generation
of laser facilities~\cite{papadopoulos,weber.mre.2017,gales.rpp.2018}.
The stronger the electromagnetic field, the lower the electron energy
that is needed to reach high $\chi_e$.
In experiments with aligned crystals where the
field strength $\sim 10^{13}~\Vm$~\cite{uggerhoj.rmp.2012}, $\chi_e \simeq 1$ and evidence of
quantum radiation reaction require 100-GeV electron beams~\cite{wistisen.nc.2018}.
Earlier experiments achieved higher $\chi_e \simeq 7$ with the use of
tungsten, rather than silicon, targets due to the stronger nuclear field~\cite{kirsebom.prl.2001}.
$\chi_e > 1$ will also be probed in beam-beam interactions in the next
generation of linear colliders~\cite{chen.prl.1989,esberg.prstab.2014}.

Despite the strong spatial and temporal compression of laser pulses,
it is inevitable that the electron will have to traverse a finite region
of space over which the field strength ramps up before it reaches the point of
peak $a_0$.
If significant energy loss takes
place during this interval, the electron's $\chi_e$ will be much smaller
than that predicted by \cref{eq:ChiMax}.
We now derive a scaling for the maximum $\chi_e$ reached by an
electron, which accounts for radiative losses and the spatial structure
of the laser pulse, following~\cite{blackburn.pra.2017}.


Consider an electron colliding at angle $\theta$ with a linearly
polarized laser pulse that has Gaussian temporal and radial intensity profiles
of size $\tau$ and $r_0$ respectively. Here $\tau$ is the full
width at half maximum (FWHM) of the temporal intensity profile
and $r_0$ is the waist of the focussed pulse (the radius at which the
intensity falls to $1/e^2$ of its peak).
As the crossing angle $\theta$ is not necessarily zero, we must take the
transverse structure of a focussed laser pulse into account. In our Monte Carlo
simulations, the spatial dependence of the electromagnetic field is treated as
a tightly focussed Gaussian beam with waist size $r_0$ and Rayleigh length
$z_R = \pi r_0^2/\lambda$. The fields are calculated up to fourth-order in the
diffraction angle $r_0/z_R$, following \cite{salamin.apb.2007}, and therefore
go beyond the paraxial approximation. Nevertheless, in order to obtain a
relatively simple analytical expression for the maximum $\chi_e$, we use
a reduced model for the fields that will, as we show, capture the essential physics.

The laser pulse is treated as a `light bullet', with Gaussian
temporal and transverse spatial profiles of constant size. We also neglect
the longitudinal components of the fields and wavefront curvature, such that
the pulse becomes a plane electromagnetic wave. For all the waist sizes under
consideration (generally at least $r_0 = 2.5 \lambda$, where $\lambda$ is
the wavelength), the transverse components provide the dominant contribution
to $\chi_e$. We assume that the electron Lorentz factor $\gamma \gg  a(\phi)$,
where $a(\phi) = e E(\phi)/(m c \omega_0)$, the local value of the normalized
electric field $E$ at phase $\phi$, at least up to the point where its
quantum parameter is maximized. This occurs before the electron has undergone
substantial energy losses, after which ponderomotive forces can eject the
electron from the laser pulse at large angle~\cite{li.prl.2015}, and our assumption
that the trajectory is ballistic breaks down.

Under these circumstances,
the envelope of the normalized electric field along the electron trajectory
is given by $a(x,y,z,t) \simeq a_0 \exp[-(x^2+y^2)/r_0^2 - 2\ln{2}(t-z)^2/\tau^2]$,
and $x = -t \sin\theta$, $y = 0$, $z = -t \cos\theta$. Here we have used the fact
that the plane in which the collision angle lies may be chosen arbitrarily.
This is written more compactly as~\cite{blackburn.pra.2017,blackburn.ppcf.2018}
	\begin{align}
	a(\phi) &= a_0 \exp[-\ln(2)\,\phi^2/(2\pi^2 \neff^2)],
	\label{eq:EffectiveA}\\
	\neff &= \frac{\omega_0 \tau}{2\pi} \left[1 + \frac{\tau^2 \tan^2(\theta/2)}{r_0^2 \ln 4}\right]^{-1/2},
	\label{eq:EffectiveN}
	\end{align}
defining the phase $\phi = (1 + \cos\theta) \omega_0 t$ and an effective duration
(per wavelength) $\neff$.
Carrier envelope phase effects may be neglected, as done here, provided $\neff \gtrsim 2$.
The point at which $\chi_e$ is maximized is defined by $[\gamma(\phi) a(\phi)]' = 0$,
where primes denote differentiation with respect to phase.

We substitute into this extremization condition: the $a(\phi)$ and $a'(\phi)$ given by
\cref{eq:EffectiveA}; and $\gamma'(\phi) = \Power / [2 (1+\cos\theta) m \omega_0]$,
where $\Power = 2\alpha m^2 \chi_e^2 g(\chi_e) / 3$ is the instantaneous
radiated power (per unit time).
The Gaunt factor $0 < g(\chi_e) \leq 1$ accounts for quantum corrections to the photon
spectrum that reduce the the radiated power from its classically predicted
value~\cite{erber.rmp.1966};
the factor of $\frac{1}{2}$ in $\gamma'(\phi)$ comes from averaging over the $\sin^2$ oscillation
of the electric field (recall that \cref{eq:EffectiveA} defines the envelope
of the field and the pulse is linearly polarized).
Then factors of $\phi$ are traded for $\chi_e$ using $\chi_e(\phi) =
\gamma(\phi) a(\phi) \omega_0 (1+\cos\theta)/ m$.
The remaining dependence on $\gamma(\phi)$ is removed by setting
$\gamma(\phi) = \gamma_0$, the electron's initial Lorentz factor.

This is motivated by the probabilistic nature of radiation losses
in the quantum regime, which means that $\chi_e(\phi)$ is not single-valued
at a given phase. Instead, there is a distribution $f(\chi_e,\phi)$ that
evolves as the electron population travels through the laser pulse.
The highest $\chi_e$ is reached by electrons that lose less energy
than would be expected classically. This phenomenon
is called `straggling'~\cite{shen.prl.1972,duclous.ppcf.2011}, or
`quenching' in pulses so short that it is possible that the electron
does not radiate at all~\cite{harvey.prl.2017}.
Such electrons are less affected by ponderomotive deflection as
their energy remains large, which supports our assumption that the
trajectory remains approximately ballistic at least up to the point
at which $\chi_e$ is maximized.
As our scaling is concerned with this part of the electron distribution
function, setting $\gamma = \gamma_0$ is a way to isolate these electrons.

We find that maximum quantum parameter $\chi_\text{max}$ satisfies
	\begin{multline}
	\frac{\chi_\text{max}^4 g^2(\chi_\text{max})}{\chimax^4} = \\
		\frac{9 \ln{2}\,(1+\cos\theta)^2}{(\pi R_c \neff)^2}
		\ln\!\left[ \frac{(1+\cos\theta) \chimax}{2 \chi_\text{max}} \right].
	\label{eq:ChiMaxScaling}
	\end{multline}
Here $\chi_0$ is the largest possible quantum parameter [\cref{eq:ChiMax} with $\theta = 0$]
and the classical radiation reaction parameter $R_c = \alpha a_0 \chimax$%
~\cite{koga.pop.2005,dipiazza.prl.2010}.

In the limit $\chi_\text{max} \ll 1$, \cref{eq:ChiMaxScaling} has a solution
in terms of the Lambert function $\mathscr{W}$, which is defined
for real $x > 0$ by $x = \mathscr{W}(x) \exp[\mathscr{W}(x)]$:
	\begin{align}
	\frac{\chi_\text{max}}{\chimax} &=
		\frac{1+\cos\theta}{2}
		e^{-\mathscr{W}(\delta^2)/4},
	\label{eq:ClassicalChiMax}
	\\
	\delta &=
		\frac{\pi R_c \neff (1+\cos\theta)}{3 \sqrt{2 \ln{2}}}.
	\end{align}
$\mathscr{W}(\delta^2)$ is strictly increasing for $\delta \geq 0$
and therefore $\chi_e$ decreases with increasing $\delta$.
Unlike \cref{eq:ChiMax}, \cref{eq:ClassicalChiMax}
does not depend symmetrically upon $a_0$ and $\gamma_0$,
as $\delta \propto a_0^2 \gamma_0$.
Hence it is more beneficial to increase $\gamma_0$
than $a_0$ when aiming for very large $\chi_e$. Physically
this is because the photon emission rate has a stronger dependence
on $a_0$ than on $\gamma_0$; by minimizing the number of emitted photons
we also mitigate the radiative losses that would reduce $\chi_e$.
Indeed, $\chi_\text{max}$ is generally smaller than $\chi_0$
because it is reached in the rising edge of the pulse,
before the electron encounters the point of highest intensity~\cite{blackburn.pra.2017}.
Compare \cref{eq:ChiMax} and \cref{eq:ClassicalChiMax}:
the scaling of $\chi_\text{max}$ with $a_0$ is much weaker in the latter
case, because peak value of $a_0$ does not contribute fully.


To show that \cref{eq:ChiMaxScaling} can be used as a
quantitative prediction of the largest $\chi_e$ that is reached in a
laser-electron beam collision, we compare its predictions to the
results of single-particle Monte Carlo simulations.
These model a QED cascade of photon emission and pair creation by
factorizing it into a product of
first-order processes~\cite{ridgers.jcp.2014,gonoskov.pre.2015},
which occur along the particles' classical trajectories
at positions determined by integration of QED probability rates
that are calculated in the locally constant field approximation~\cite{ritus}.
This `semiclassical' approach is valid when $a_0^3/\chi_e \gg 1$
because the formation lengths of the photons and electron-positron pairs are
much smaller than the laser wavelength and interference effects are
suppressed~\cite{dinu.prl.2016}.
Details of the simulations are given in \cref{sec:Simulations}.

	\begin{figure}
	\centering
	\includegraphics[width=\stdwidth]{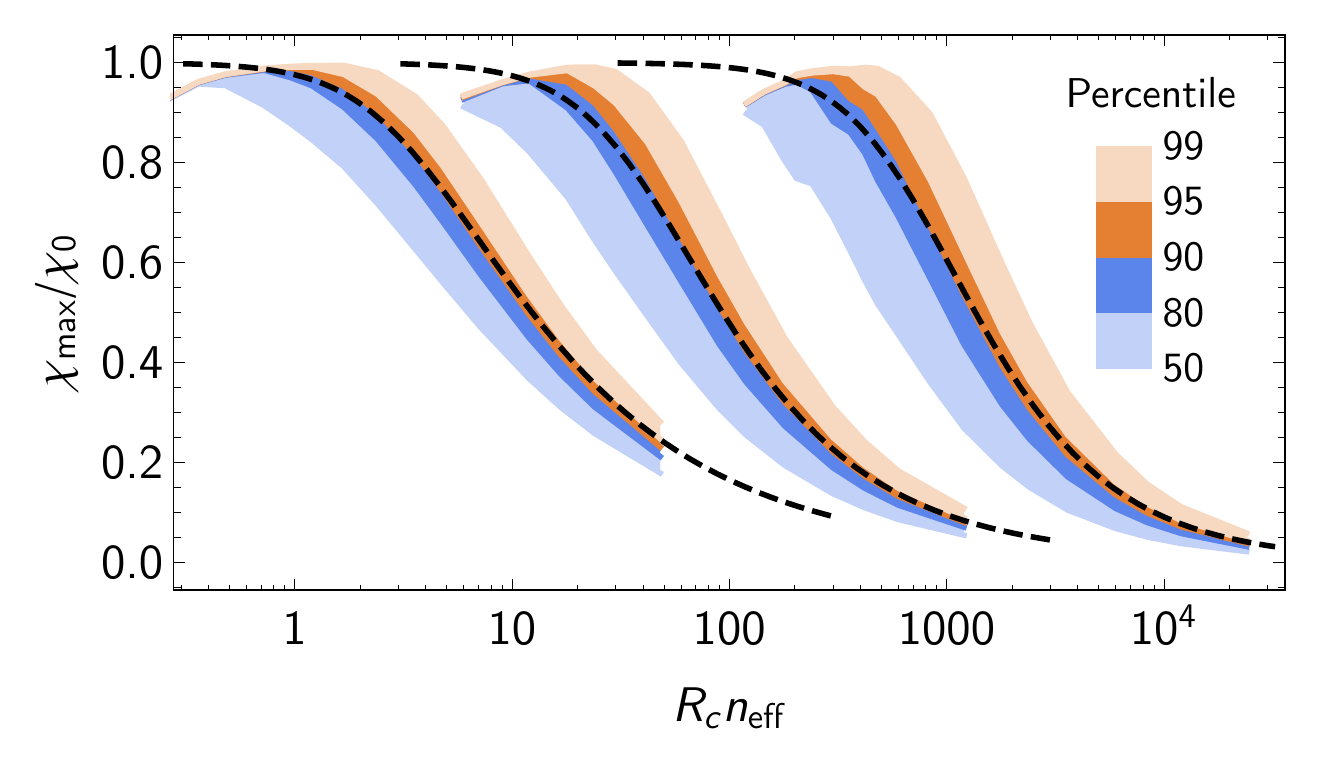}
	\caption[Distribution of the maximum $\chi$]
		{Radiation reaction limits the maximum quantum parameter
		reached by the electron $\chi_\text{max}$:
		the distribution of $\chi_\text{max}$ reached in a
		laser-electron collision where the largest possible $\chi_e = \chimax$
		is (colour scale, left to right) 1, 10 and 100
		from simulations and (dashed lines) our analytical prediction
		of the same, \cref{eq:ChiMaxScaling}.
		See text for other collision parameters.
		}
	\label{fig:ChiPercentiles}
	\end{figure}
	
Starting with head-on collisions, we show how the distribution of
$\chi_\text{max}$, the largest quantum parameter attained by the
electron on its passage through the laser pulse,
is affected by the duration of a linearly polarized, plane-wave laser pulse.
The electron initial Lorentz factor $\gamma_0$ is set to be one of
$5\times10^3$, $2\times10^4$, and $10^5$.
The laser $a_0$ is chosen such that $\chi_0$ is 1, 10 and 100 respectively.
The laser frequency is fixed at $\omega_0 = 1.55$~eV and the pulse duration $\tau$
is varied from 2 to 200 wavelengths. The distributions of $\chi_\text{max}$
shown in \cref{fig:ChiPercentiles}
demonstrate that increasing the pulse duration strongly reduces
the number of electrons that reach large quantum parameter. This
behaviour is captured by \cref{eq:ChiMaxScaling}, which we find to be
a good quantitative prediction of the 90th percentile of the distribution,
provided $\neff \gtrsim 2$. Otherwise the specific value of the carrier
envelope phase $\phi_\text{CEP}$ must be taken into account~\cite{mackenroth.prl.2011},
as the maximal electric field of a pulse $E(\phi) \propto a(\phi) \sin(\phi + \phi_\text{CEP})$
is smaller for $\phi_\text{CEP} = 0$ than $\phi_\text{CEP} = \pi/2$,
and the difference grows as the duration is reduced.
We set $\phi_0 = 0$ throughout this paper, which is why the upper bounds of
the distributions shown in \cref{fig:ChiPercentiles} roll over as $R_c \neff$ is reduced.
They would saturate at $\chi_\text{max} = \chimax$ if instead $\phi_0 = \pi/2$.

\Cref{eq:ChiMaxScaling} can be solved to find the largest laser pulse duration
$\tau$ for which $\chi_\text{max} > \chi_0/2$.
We find that $\tau \lesssim 8 m \gamma_0 / \mathcal{P}(\chi_0/2)$,
where $\mathcal{P}$ the radiated power (including
quantum corrections) of an electron with given $\chi$.
The larger the radiated power, the shorter
the pulse must be to ensure that at least 10\% of the electrons
reach a quantum parameter of at least $\chi_0/2$.
For the three cases shown in \cref{fig:ChiPercentiles}, we predict
the duration $\tau$ can be at most 137, 41.2 and 30.0~fs before
this happens, in good agreement with the simulation results
where the largest $\tau$ is $131$, $41.7$ and $30.1$~fs respectively.

\section{Enhanced signatures of quantum effects}

	\begin{figure}
	\centering
	\includegraphics[width=\stdwidth]{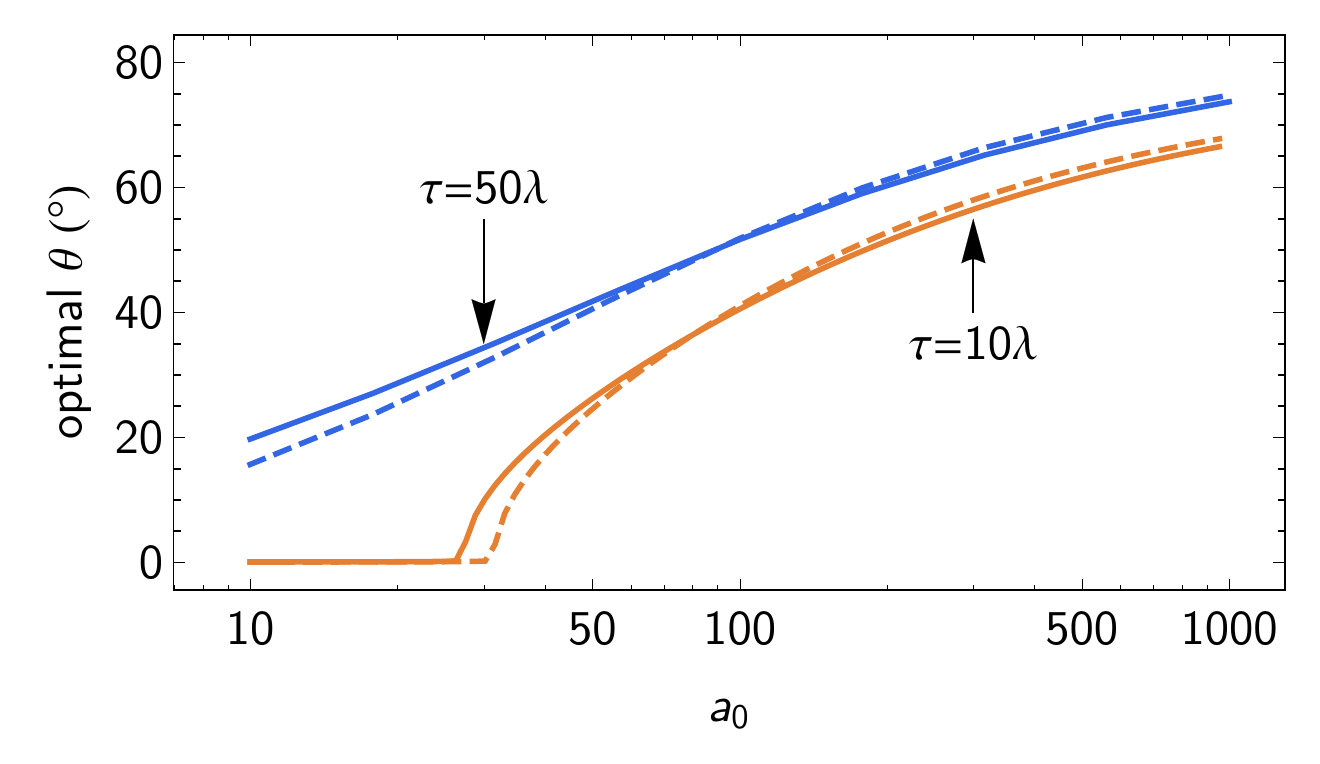}
	\caption[Optimal angle]
		{The angle at which $\chi_\text{max}$ is largest,
		as predicted by \cref{eq:ChiMaxScaling}: for a
		collision between an electron beam with $\gamma_0 = 2\times10^4$
		(solid) and $5\times10^3$ (dashed) and a laser pulse with
		given $a_0$, wavelength $\lambda = 0.8~\micron$, spot size
		$r_0 = 2~\micron$ and duration
		$\tau = 50\lambda$ (blue) and $10\lambda$ (orange).
		}
	\label{fig:OptimalAngle}
	\end{figure}

\Cref{eq:ChiMax} leads us to expect that quantum effects are strongest
in the head-on collision geometry, where the geometric factor
$1+\cos\theta$ is largest.
However, unless the pulse duration is as little as
a few cycles in length (when radiation `quenching' is possible%
~\cite{harvey.prl.2017}), radiation reaction strongly reduces the number
of electrons that get close to the maximum possible $\chi_e$.
This can be mitigated by moving to collisions at oblique incidence,
because the spot to which a laser pulse is focussed
($\sim 2~\micron$) is typically smaller than the length of its temporal profile
(20~fs~\cite{papadopoulos}, 30~fs~\cite{bahk,pirozhkov,gales.rpp.2018} or 150~fs~\cite{weber.mre.2017}).
Even though the maximum possible $\chi_e$ at $\theta > 0$ is smaller than
that at $\theta = 0$, many more electrons get close to the maximum
because the interaction length is shorter and radiative
losses are reduced.
This is illustrated in \cref{fig:OptimalAngle}, where the collision
angle $\theta$, predicted by \cref{eq:ChiMaxScaling} to maximize
$\chi_\text{max}$, is plotted for
two exemplary pulse durations $\tau = 10\lambda$ and $50\lambda$
(27 and 130~fs respectively at a wavelength of $0.8~\micron$).
The shorter the pulse duration, the larger $a_0$
can be before the head-on collision ceases to be optimal.
As the laser amplitude is increased, radiation reaction becomes
stronger and the optimal angle increases away from zero.
The increased $\chi_\text{max}$ at oblique incidence enhances
two quantum effects: the emission of photons with
energy comparable to that of the electron,
and the stochastic broadening of the electron energy spectrum.

	\begin{figure}
	\centering
	\includegraphics[width=\stdwidth]{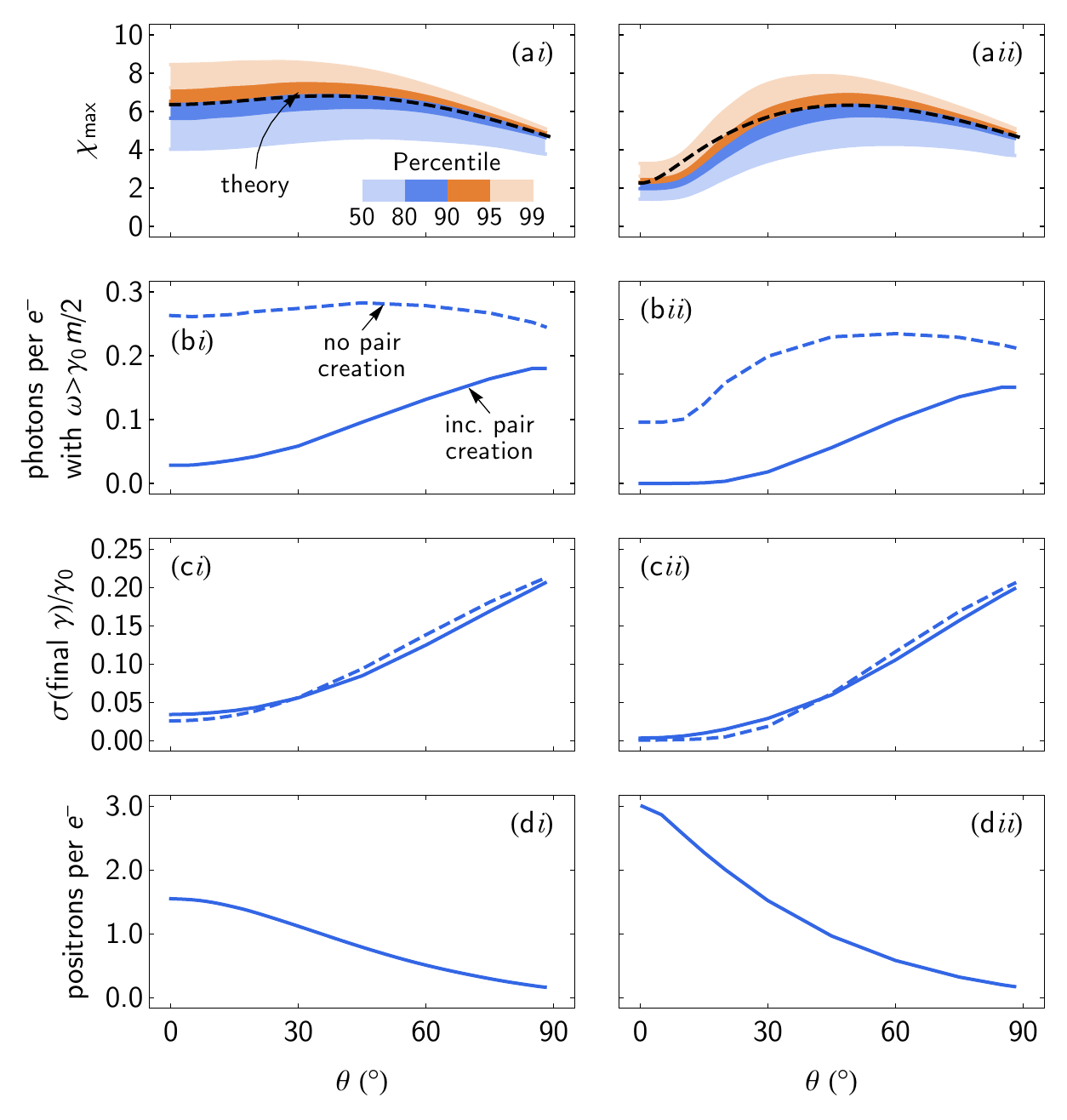}
	\caption[Quantum effects at oblique incidence]
		{Enhanced quantum effects at oblique incidence:
		(a) distribution of $\chi_\text{max}$;
		(b) the number of photons per electron with energy
		$\omega > \gamma_0 m / 2$;
		(c) the standard deviation of the final
		$\gamma$;
		and (d) the number of positrons per electron,
		after electrons with $\gamma_0 = 2\times10^4$
		collide at angle $\theta$ with a laser pulse with
		$a_0 = 82.4$, wavelength $\lambda = 0.8~\micron$,
		focal spot size $r_0 = 2~\micron$ and a duration
		of (i) $\tau = 10\lambda$ and (ii) $50\lambda$.
		In (b,c) results are from simulations with (solid)
		and without (dashed) electron-positron pair creation.
		}
	\label{fig:CollisionAngle}
	\end{figure}

In \cref{fig:CollisionAngle} we show how these two signatures are affected
by the collision angle $\theta$ in a QED cascade when $\chimax = 10$ and the laser pulse
duration is one of $\tau = 10\lambda$ and $50\lambda$.
As the (linearly polarized) laser is focussed tightly to a spot size of $r_0 = 2~\micron$,
the electromagnetic field in our simulations is calculated up
to fourth-order in the diffraction angle $\epsilon = r_0 / z_R$ where
$z_R = \pi r_0^2/\lambda$ is the Rayleigh range~\cite{salamin.apb.2007}.
Further details of the simulations are given in \cref{sec:Simulations}.
The dependence of the distribution of $\chi_\text{max}$
on the angle is different in the two cases: whereas it is approximately
constant at $\chi_\text{max} \simeq 5$ for the shorter pulse, the
maximum quantum parameter is strongly suppressed for $\theta < 15\degree$
for the longer pulse. We find that not only is $\chi_\text{max}$ maximized
at $\theta \simeq 45\degree$ rather than at $0\degree$, as shown in
\cref{fig:OptimalAngle}, but that the
value at $90\degree$ is twice that at $0\degree$.
The reduced apparent pulse duration at normal incidence
more than compensates for the reduction in the geometric factor in $\chi_e$
because it reduces the electron's total loss of energy to radiation.
Our theoretical scaling \cref{eq:ChiMaxScaling} captures both these
effects, in close agreement with the simulation results.

The number of high-energy photons is especially sensitive to the
highest $\chi_e$ reached by the electron~\cite{duclous.ppcf.2011,blackburn.prl.2014}.
Accordingly, consider the number of photons $N_\gamma$ with energy
$\omega > \gamma_0 m / 2$ in the absence of electron-positron pair creation
[the dashed lines in \cref{fig:CollisionAngle}(b)]. For the shorter pulse,
$N_\gamma$ is almost independent of the collision angle, whereas for the longer
pulse, it is maximized at $\theta \simeq 45\degree$ and suppressed
for $\theta < 15\degree$~\cite{blackburn.ppcf.2015}.
In both cases the dependence of $N_\gamma$ on $\theta$ mimics that of $\chi_\text{max}$.
When depletion of the photon spectrum due to electron-positron pair creation
is included,
the optimal angle is increased to $90\degree$ for both pulse durations.
This is because the reduced interaction length at normal incidence suppresses the pair
creation probability~\cite{blackburn.pra.2017}, as shown in \cref{fig:CollisionAngle}(d).

Another important signature of quantum effects on radiation reaction
is broadening of the electron energy spectrum~\cite{neitz.prl.2013},
caused by the stochasticity of photon emission~\cite{shen.prl.1972}.
The variance of the energy distribution $\sigma_\gamma^2$ is
studied in detail in \cite{yoffe.njp.2015,vranic.njp.2016,ridgers.jpp.2017,niel.pre.2018},
where it is shown that the temporal evolution of the variance is
governed by two competing terms: one that arises because
the radiated power is larger for higher energy electrons,
which favours decreasing $\sigma_\gamma$,
and a stochastic term, which favours increasing $\sigma_\gamma$.
The broadening term dominates if $\chi_e$ is
large and the pulse duration is short.
Both of these cause oblique incidence to be favoured for the scenario
explored in \cref{fig:CollisionAngle}: $\chi_\text{max}$ is
larger at $\theta > 0$ (or at least, not significantly reduced) and
the interaction length is shorter as well.
\Cref{fig:CollisionAngle}(c) shows that the variance of the post-collision
energy is larger for larger $\theta$~\cite{yoffe.spie.2017},
and that this is not changed appreciably by electron-positron pair creation.

\section{Towards radiative corrections in strong-field QED}

We now consider the collision parameters necessary to reach
$\alpha \chi_e^{2/3} \gtrsim 1$, where strong-field QED
is conjectured to become fully nonperturbative.
By this we mean that perturbation theory with respect to the
dynamical electromagnetic field breaks down~\cite{narozhny.prd.1980}:
for example, the lowest order correction to the strong-field QED
vertex $V^{(1)}_\mu = i e \gamma_\mu$ grows as
$V^{(3)} \sim \alpha \chi_e^{2/3} V^{(1)}$~\cite{morozov.jetp.1981}
(debated in \cite{gusynin.prb.1999}).
Recall that the theory is already nonperturbative in the sense
that amplitudes must be calculated to all orders in coupling to the background
electromagnetic field $a_0$ if $a_0 > 1$~\cite{ritus}.
The qualitative difference from perturbative QED is that radiative
corrections are expected to grow as power laws, rather than
logarithmically, in the strong-field regime~\cite{fedotov.jpc.2017}
(the transition between the two is studied in \cite{podszus.prd.2019,ilderton.prd.2019}).

Reaching such large $\chi_e$ is therefore of fundamental interest,
but experimentally challenging.
The obstacle is severe radiation losses at large $\chi_e$:
in the case where the strong field is provided by a tightly focussed,
ultraintense laser, we show
how the collision angle plays an important role in mitigating these
losses by reducing the interaction time.
In the beam-beam geometry Yakimenko~\textit{et al.}~\cite{yakimenko.arxiv.2018} propose to
reach $\alpha \chi_e^{2/3} \sim 1$, the electron-bunch length plays the important role;
in an alternative geometry of laser--electron-beam collision proposed by
Baumann~\textit{et al.}~\cite{baumann.arxiv.2018},
the interaction time is reduced by plasma-based compression of a single-cycle laser pulse
to sub-femtosecond duration, in advance of the collision.
Strictly the calculation cannot be done for $\alpha \chi_e^{2/3} \sim 1$,
because we would need all the radiative corrections; however, we
can estimate when they become significant by using our results to
find the collision parameters necessary to reach, say, $\chi_e = 100$,
at which the vertex correction is of order $10\%$
and radiative corrections begin to become non-negligible.
Note that the energy loss which reduces $\chi_\text{max}$ from
$\chi_0$ occurs within the intensity ramp where radiative corrections
are less important. Hence, while our analysis neglects such corrections,
the crucial physical insight remains accurate.

	\begin{figure}
	\centering
	\includegraphics[width=0.9\linewidth]{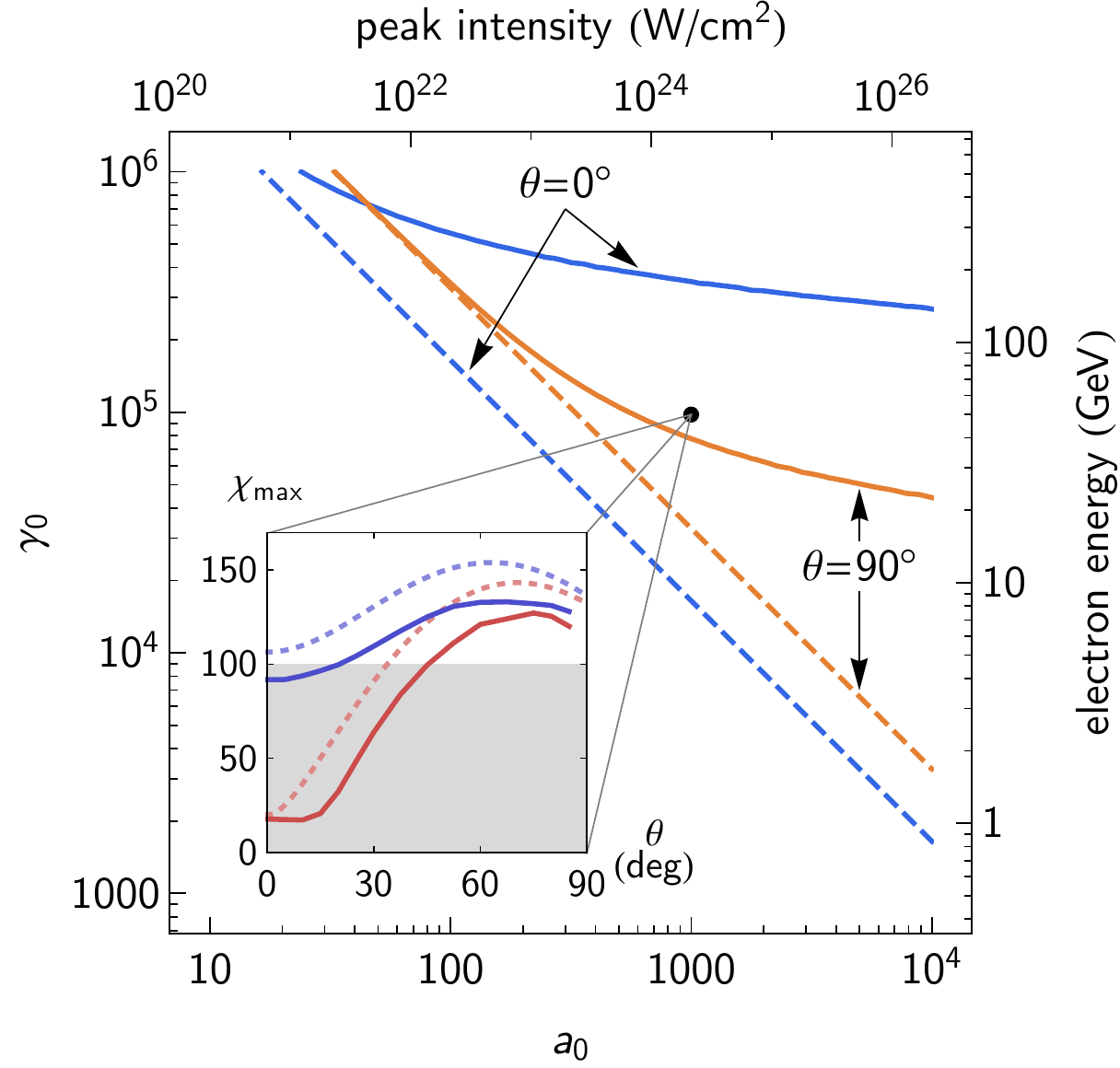}
	\caption[Breakdown of perturbative strong-field QED]
		{
		The minimum $\gamma_0$ and $a_0$ required for
		$\chi_e \geq 100$ ($\alpha \chi_e^{2/3} \geq 0.16$):
		we compare \cref{eq:ChiMax} (dashed lines), which
		neglects radiative losses, with \cref{eq:ChiMaxScaling},
		which includes them, for an electron colliding with
		a linearly polarized laser pulse (wavelength $\lambda = 0.8~\micron$,
		duration $\tau = 50\lambda$ and focal spot size
		$r_0 = 2~\micron$).
		We see that collisions at normal incidence (orange)
		are very strongly favoured over head-on (blue)
		when radiative losses are accounted for.
		(inset) Theoretical $\chi_\text{max}$ \cref{eq:ChiMaxScaling} (dashed)
		and the 90th percentile of the $\chi_\text{max}$ distribution
		from simulations (solid)
		as a function of angle $\theta$ for a collision at
		50~GeV and $a_0 = 1000$ for pulse duration of
		$10\lambda$ (purple) and $50\lambda$ (red).
		}
	\label{fig:Breakdown}
	\end{figure}

The dashed lines in \cref{fig:Breakdown} show the minimum $\gamma_0$
and $a_0$ if $\chi_e$ were given by \cref{eq:ChiMax}: it is evident that
the ideal collision angle $\theta = 0$.
This is no longer the case when dynamical effects are taken into account.
Using \cref{eq:ChiMaxScaling} to estimate the minimum energy
and laser intensity instead, we find that collisions at $\theta = \pi/2$
are strongly favoured for a pulse with duration $\tau = 50\lambda$.
The additional electron energy or laser intensity necessary to compensate
radiative losses can be substantial, which is indicated by the vertical
(horizontal) gaps between the solid and dashed lines.
At $a_0 = 1000$ and $\theta = 0$, for example, the minimum energy must be
increased by more than an order of magnitude, from the naive estimate of 8.4~GeV to 180~GeV.
The gradient of the lines indicates that the necessary increase in
$\gamma_0$ is always smaller than the equivalent increase in $a_0$.
As discussed earlier, this is because of the stronger dependence of the
photon emission rate on $a_0$.

At $a_0 = 1000$, which is equivalent to an intensity of $2\times10^{24}~\Wcmsqd$
at a wavelength of $0.8~\micron$, the energy required to reach $\chi_e = 100$
and the onset of radiative corrections is ${\sim}40$~GeV,
at which point oblique incidence is favoured for both $\tau = 10\lambda$
and $50\lambda$ (see inset of \cref{fig:Breakdown}).
This energy is readily achievable with conventional
accelerators~\cite{bula.prl.1996,burke.prl.1997} and the necessary
laser system is of the kind being commissioned at the ELI facilities~\cite{eli.tdr}.
The required laser intensity may be reduced at the expense of increasing
the electron beam energy;
according to \cref{fig:OptimalAngle}, this reduces the optimal angle of incidence,
whereas tighter focussing, i.e. reduction of $r_0$, would cause it to increase.
It is important to note that the laser intensity cannot be reduced to an
arbitrarily low level, and the electron energy increased to compensate,
because power-law growth of the radiative corrections
occurs only in the high-intensity limit $a_0 \gg 1$;
if $\alpha \chi_e^{2/3} \gtrsim 0.1$ is approached by means of ever higher
electron energies instead, that growth would be logarithmic as in perturbative
QED~\cite{podszus.prd.2019,ilderton.prd.2019}.

$\chi_\text{max}$ increases as $r_0$ becomes smaller, assuming oblique incidence
and fixed $a_0$. Tighter focussing is therefore motivated, not only to achieve the
highest possible intensity, but also to reduce radiative energy losses.
This can also be interpreted as a minimal requirement on the
quality of the focussing. `Wings' around the focal spot would increase the interaction
time, which effectively increases the spot size $r_0$ in \cref{eq:ChiMaxScaling}.
Consider, for example, the collision of a 50-GeV electron beam with a pulse that has
radial profile $a(r) = a_0 [(1-\delta) e^{-r^2/r_0^2} + \delta e^{-r^2/(f r_0)^2}]$
at oblique incidence $\theta = 85\degree$.
We set $\delta = 0.1$, $r_0 = 2~\micron$ and increase $f$ from $1$ to $2$, causing
a shoulder to develop in the intensity profile at the focal plane.
The increased interaction time increases the energy loss of the electron beam
and reduces the maximum $\chi_e$ reached: simulations indicate that at
$\chimax = 1000$, the 90th percentile of $\chi_\text{max}$ is reduced by 15\%,
from $0.13\chimax$ to $0.11\chimax$.
Increasing $r_0$ from 2 to $2.55~\micron$ would, according to \cref{eq:ChiMaxScaling},
cause the same decrease and therefore the latter may be regarded as an \emph{effective}
focal spot size for the modified radial profile.
An extension of \cref{eq:ChiMaxScaling} for more general radial and temporal
intensity profiles will be addressed in future work.

Alignment of the beams is, admittedly, more challenging at oblique incidence than it is
for head-on collisions, which has been the focus of previous experimental
work on radiation reaction~\cite{cole.prx.2018,poder.prx.2018}.
Nevertheless, an argument in its favour that the initial beam and its collision products
are directed well away from the laser focussing optic.
Furthermore, if the laser pulse is sufficiently intense or long that
radiation reaction would suppress $\chi_\text{max}$ well below the $\chi_e$
necessary to observe the onset of radiative corrections, then regardless
of whether the beams overlap or not, the collision will be unsuccessful
in reaching the regime in question.
Our results, including \cref{eq:ChiMaxScaling}, can be used to determine whether
it is possible for a particular set of collision parameters.
Successful overlap between the beams can be identified
by means of coincidence measurements of the $\gamma$-ray flash, the
electron energy loss and positron production,
because as \cref{fig:CollisionAngle} shows, the numbers of high-energy photons
and positrons are sensitive to the highest $\chi_e$ reached and the duration
over which it is sustained.

\section{Summary}

We have studied how to reach large quantum parameter in the collision
of an electron beam with an intense laser pulse.
Our scaling for the maximum $\chi_e$, which is verified by
Monte Carlo simulations, predicts that the optimal collision geometry
is not head-on for long or high-intensity laser pulses.
This is because of strong radiative losses, which reduce the
electron energy and so its quantum parameter.
The growth of $\chi_e$ is then much weaker than the linear
scaling of the naive prediction, which ignores radiation losses and thereby
overestimates the efficiency of $\chi$-generation.
The shorter interaction length at oblique incidence
compensates for the geometric reduction in $\chi_e$, causing
signatures of quantum effects to be enhanced at $\chi_e = 10$.
Beyond their applicability to nearer term experiments,
our results show
that a collision at oblique incidence is a viable platform for studying
the onset of the breakdown of perturbative strong-field QED at
$\alpha \chi_e^{2/3} \gtrsim 0.1$.
It is be to hoped that the feasibility of reaching
this regime in a future high-intensity laser experiment will
further motivate the theoretical work necessary to identify its
explicit signatures, and how modifications to the photon emission
and pair creation rates manifest themselves.

\begin{acknowledgments}
The authors acknowledge support from the Knut and Alice
Wallenberg Foundation (T.G.B. and M.M.), the Swedish Research Council
(grants 2013-4248 and 2016-03329, M.M.)
and the Engineering and Physical Sciences Research Council
(grant EP/M018156/1, C.P.R.).
Simulations were performed on resources provided by the Swedish National
Infrastructure for Computing (SNIC) at the High Performance Computing
Centre North (HPC2N).
\end{acknowledgments}

\appendix
\section{Monte Carlo simulations}
\label{sec:Simulations}

In this appendix, we summarize the method by which the interaction
between electrons and intense lasers can be modelled in the quantum
radiation reaction regime.

In the semiclassical approach to the collision process, the electron
follows a (radiation-free) classical trajectory between point-like,
probabilistically determined, QED events. These events are implemented
using the standard Monte Carlo algorithm~\cite{ridgers.jcp.2014,gonoskov.pre.2015},
with rates calculated in the locally constant field approximation~\cite{erber.rmp.1966,ritus}.
We use \texttt{circe}, a particle-tracking code that simulates photon and
positron production by high-energy electrons and photons in intense laser pulses.
Collective effects and back-reaction on the external field are neglected,
as appropriate for the charge densities under consideration here.
In between emissions, the particle trajectories follow from the Lorentz force equation.
If the external field is a plane wave, the particle push takes the following
form~\cite{blackburn.pop.2018}:
the spatial components of the momentum $p^\mu$ perpendicular to the laser
wavevector $k$ are determined by
$\partial_\phi \vec{p}_\perp = -e \vec{E}_\perp(\phi)/\omega_0$,
where $\vec{E}_\perp$ is the electric field at phase $\phi$
and the angular frequency $\omega_0 = k^0$.
The other two components follow from the conditions $k.p = \text{const}$
and $p^2 = m^2$, and the position from $\partial_\phi x^\mu = p^\mu / k.p$.
If the field is a focussed Gaussian beam, and therefore a function of all
three spatial coordinates, we use the particle push introduced by Vay~\cite{vay.pop.2018}
and the analytical expressions given in~\cite{salamin.apb.2007}.

To model photon emission, each electron is initialized with an
optical depth against emission $\tau = -\log(1-R)$ for pseudorandom
$0 \leq R < 1$, which evolves as $\partial_t \tau = W(\chi_e,\gamma)$,
where $W(\chi_e,\gamma)$ is the instantaneous probability rate of emission,
$\chi_e$ the electron quantum parameter and $\gamma$ its Lorentz factor,
until the point where $\tau$ falls below zero.
Then the energy of the photon is pseudorandomly sampled from the
differential spectrum and $\tau$ is reset.
We assume that emission occurs in the direction parallel to the initial
momentum, as the electron emits into a narrow cone of opening angle
$1/\gamma$, which determines the electron momentum after the scattering
by the conservation of momentum.
The most stringent restriction on the timestep $\Delta t$ at high intensity
is that the probability of multiple emissions per step be
much smaller than 1, i.e. $\Delta\tau \ll 1$.
The timestep is then determined by $\Delta\tau \simeq 1.44\alpha \chi_0 \Delta t/\gamma_0 \leq 10^{-2}$,
or $\omega_0 \Delta t / (2\pi) \leq 10^{-2}$, whichever leads to the smaller result.
Electron-positron pair creation by photons is modelled in an analogous
way, except the photons follow a ballistic trajectory from their
point of creation, and on the creation of the pair, the parent photon
is deleted from the simulation.
At least $10^6$ electrons are used per simulation to generate sufficient
statistics.

\bibliography{references}

\end{document}